\documentclass[prl,aps,superscriptaddress,footinbib,10pt,twocolumn]{revtex4-2}
\usepackage[utf8]{inputenc}
\usepackage{color}
\usepackage{amsmath}
\usepackage{amssymb}
\usepackage{stmaryrd}
\usepackage{graphicx}
\usepackage[unicode=true,
bookmarks=true,bookmarksnumbered=false,bookmarksopen=false,breaklinks=false,pdfborder={0 0 1},backref=false,colorlinks=true]
{hyperref}
\hypersetup{linkcolor=magenta, urlcolor=blue, citecolor=blue, pdfstartview={FitH}, hyperfootnotes=false, unicode=true}

\makeatletter

\usepackage{xcolor}  
\usepackage{amsfonts}
\usepackage{tabularx}
\usepackage{dcolumn}
\usepackage{bm}
\usepackage{graphicx}
\usepackage{epstopdf}
\hypersetup{urlcolor=blue}

\def\ket#1{\left|#1\right\rangle}

\def\braket#1{\left\langle#1\right\rangle}

\usepackage[normalem]{ulem}

\newif\ifcmnt
\cmnttrue
\ifdefined\cmntsoff\cmntfalse\fi
\ifcmnt
    \providecommand{\aucmnt}[1]{#1}

\else
    \providecommand{\aucmnt}[1]{}

\fi

\usepackage{enumitem}
\usepackage{times}

\makeatother
\begin{document}
\title{Dynamical freezing and enhanced magnetometry in an interacting spin ensemble}

\author{Ya-Nan Lu}
\thanks{These authors contributed equally to this work.}
\affiliation{Center for Quantum Information, IIIS, Tsinghua University, Beijing 100084, China}
\affiliation{Hefei National Laboratory, Hefei 230088,  China}

\author{Dong Yuan}
\thanks{These authors contributed equally to this work.}
\affiliation{Center for Quantum Information, IIIS, Tsinghua University, Beijing 100084,  China}
\affiliation{JILA, University of Colorado Boulder, Boulder, Colorado 80309, USA}

\author{Yixuan Ma} 
\affiliation{Center for Quantum Information, IIIS, Tsinghua University, Beijing 100084,  China}
\affiliation{Hefei National Laboratory, Hefei 230088,  China} 

\author{Yan-Qing Liu}
\affiliation{Center for Quantum Information, IIIS, Tsinghua University, Beijing 100084,  China}
\affiliation{Hefei National Laboratory, Hefei 230088,  China}

\author{Si Jiang} 
\affiliation{Center for Quantum Information, IIIS, Tsinghua University, Beijing 100084,  China}
\affiliation{Hefei National Laboratory, Hefei 230088,  China}

\author{Xiang-Qian Meng} 
\affiliation{Center for Quantum Information, IIIS, Tsinghua University, Beijing 100084,  China}
\affiliation{Hefei National Laboratory, Hefei 230088, China}

\author{Yi-Jie Xu} 
\affiliation{Center for Quantum Information, IIIS, Tsinghua University, Beijing 100084,  China}
\affiliation{Hefei National Laboratory, Hefei 230088,  China}

\author{Xiu-Ying Chang} 
\affiliation{Center for Quantum Information, IIIS, Tsinghua University, Beijing 100084, China}
\affiliation{Hefei National Laboratory, Hefei 230088, China}

\author{Chong Zu} 
\affiliation{Department of Physics, Washington University, St. Louis,  Missouri 63130, USA}

\author{Hong-Zheng Zhao} 
\affiliation{School of Physics, Peking University, Beijing 100871,  China}

\author{Dong-Ling Deng}
\email[]{dldeng@tsinghua.edu.cn}
\affiliation{Center for Quantum Information, IIIS, Tsinghua University, Beijing 100084, China}
\affiliation{Shanghai Qi Zhi Institute, Shanghai 200232, China}
\affiliation{Hefei National Laboratory, Hefei 230088,  China}

\author{Lu-Ming Duan}
\email[]{lmduan@tsinghua.edu.cn}
\affiliation{Center for Quantum Information, IIIS, Tsinghua University, Beijing 100084, China}
\affiliation{Hefei National Laboratory, Hefei 230088, China}

\author{Pan-Yu Hou}
\email[]{houpanyu@tsinghua.edu.cn}
\affiliation{Center for Quantum Information, IIIS, Tsinghua University, Beijing 100084,  China}
\affiliation{Hefei National Laboratory, Hefei 230088, China}




\begin{abstract} 
\noindent 
Understanding and controlling non-equilibrium dynamics in quantum many-body systems is a fundamental challenge in modern physics \cite{Rigol2008Thermalization,Nandkishore2015many,Abanin2019Colloquium,Heyl2018Dynamical,Serbyn2021quantum,Zaletel2023Colloquium}, with profound implications for advancing quantum technologies. Typically, periodically driven systems in the absence of conservation laws thermalize to a featureless ``infinite-temperature'' state, erasing all memory of their initial conditions \cite{Lazarides2014Periodic,Lazarides2014Equilibrium,Alessio2014Long}. However, this paradigm can break down through mechanisms such as integrability \cite{Sutherland2004beautiful}, many-body localization \cite{Nandkishore2015many,Abanin2019Colloquium,Ponte2015ManyBody,Lazarides2015Fate}, quantum many-body scars~\cite{Serbyn2021quantum}, and Hilbert space fragmentation \cite{Sala2020Ergodicity,Adler2024Observation}. Here, we report the experimental observation of dynamical freezing, a distinct mechanism of thermalization breakdown in driven systems~\cite{Haldar2021Dynamical,Haldar2024Dynamical,Guo2025Dynamical,Mukherjee2024Floquet}, and demonstrate its application in quantum sensing using an ensemble of approximately $10^4$ interacting nitrogen-vacancy spins in diamond. By precisely controlling the driving frequency and detuning, we observe emergent long-lived spin magnetization and coherent oscillatory micromotions, persisting over timescales exceeding the interaction-limited coherence time ($T_2$) by more than an order of magnitude. Leveraging these unconventional dynamics, we develop a dynamical-freezing-enhanced ac magnetometry that extends optimal sensing times far beyond $T_2$, outperforming conventional dynamical decoupling magnetometry with a 4.3\,dB sensitivity enhancement. Our results not only provide clear experimental observation of dynamical freezing---a peculiar mechanism defying thermalization through emergent conservation laws---but also establish a robust control method generally applicable to diverse physical platforms, with broad implications in quantum metrology and beyond.
\end{abstract}

\maketitle

\vspace{.5cm}


\begin{figure*}[t]
\includegraphics[width=1.0\linewidth]{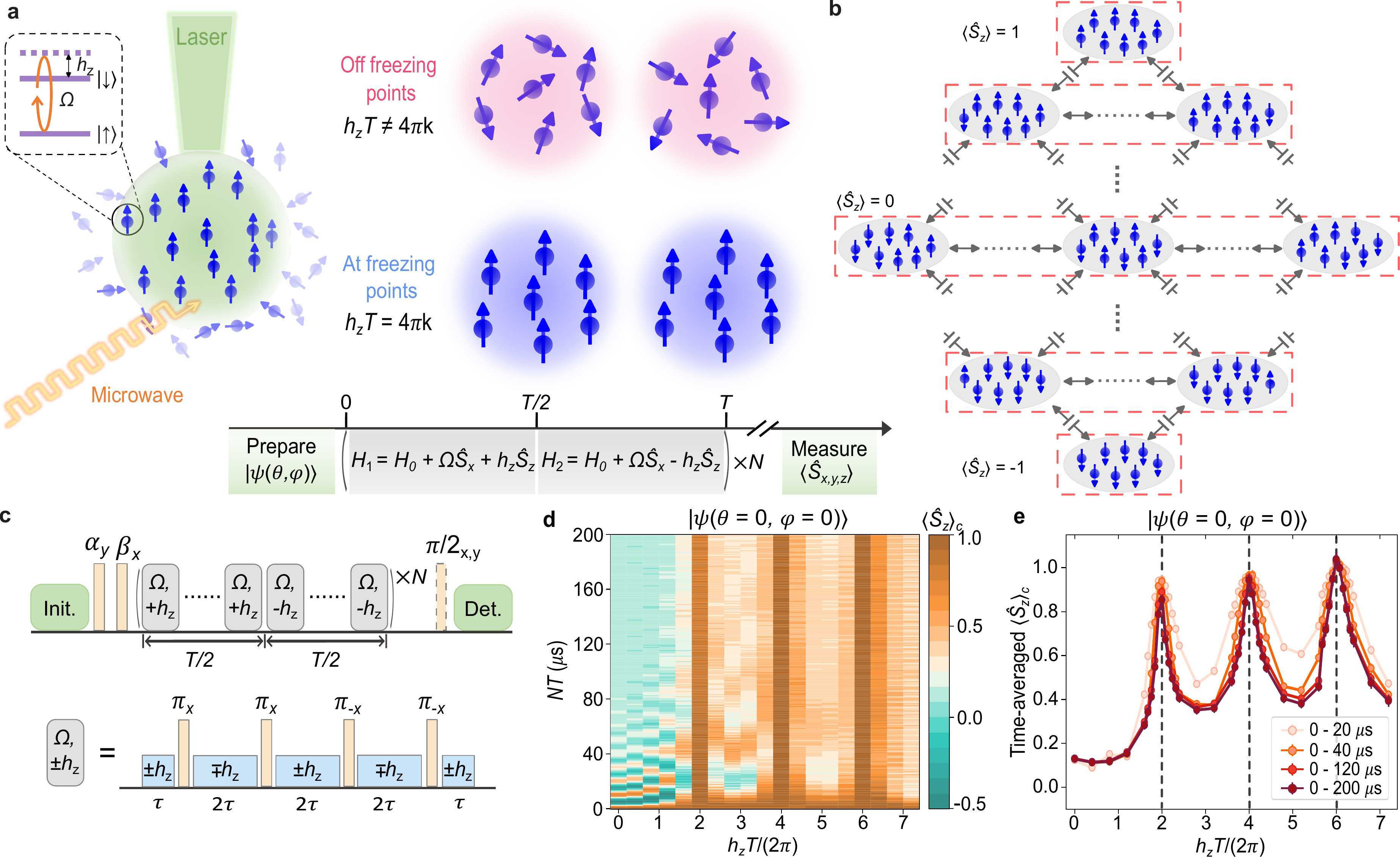}
\caption{\textbf{Dynamical freezing in an interacting spin ensemble. 
} 
\textbf{a}, Illustration of dynamical freezing. Electron spins (blue) associated with nitrogen-vacancy centers are randomly positioned in a diamond. These spins are initialized and read out via green laser illumination (green) and manipulated by global microwave fields (orange waves) with Rabi frequency $\Omega$ and detuning $h_z$ (inset).  
Experiments are implemented by initializing all spins in $\left|\psi(\theta,\varphi)\right\rangle=\cos (\theta/2)\ket{\uparrow}+e^{i\varphi}\sin(\theta/2)\ket{\downarrow}$ (e.g. $\ket{\psi}=\ket{\uparrow}$ depicted in the green blob). Then a periodic Floquet driving is implemented by alternating between two Hamiltonians, $H_1$ and $H_2$ [left and right grey boxes, $H_0$ is the static Hamiltonian shown in Eq.~\eqref{Eq:System_Hamiltonian}] with opposite detunings, every $T/2$ for $N$ cycles. The total spin magnetization $\langle \hat{S}_{\mu}\rangle$ ($\mu\in \{x,y,z\}$) is measured afterwards. At the freezing points, $\hat{\braket{S_z}}$ remains stroboscopically conserved (blue shaded regions), while away from these points it rapidly relaxes to its thermal value (pink shaded regions). 
%
\textbf{b}, Under the freezing condition, the Floquet driving dynamically conserves $\hat{\braket{S_z}}$, fragmenting the Hilbert space into disconnected sectors (dashed red boxes).
\textbf{c}, Experimental pulse sequences. Green boxes: green laser pulses for state initialization and readout. Orange boxes: strong microwave pulses with rotation angles and phases labeled at the top. Grey boxes: Floquet driving blocks consisting of strong periodic $\pi$-pulses and microwave pulses with Rabi frequency $\Omega$ (blue boxes with detuning shown inside). 
\textbf{d}, Experimental spin dynamics with the initial state $\left|\psi(0,0)\right\rangle$. Color map shows $\langle\hat{S}_{z}(t)\rangle_c$ versus $h_z$ and  $NT$ (Methods). Each data point represents the mean over $2\times10^4$ experimental trials. 
\textbf{e}, Time-averaged $\langle\hat{S}_{z}(t)\rangle_c$ obtained from the data in \textbf{d} and plotted against $h_z T$ for various integration times. All curves exhibit distinct peaks at freezing points $h_z T / (2\pi) =2,4,6$ (dashed vertical lines). Error bars represent one standard deviation. 
}\label{fig:fig1}
\end{figure*}

Quantum thermalization is a ubiquitous yet profound phenomenon~\cite{Deutsch1991Quantum,Srednicki1994Chaos}, forming the cornerstone of quantum statistical mechanics and governing diverse systems from complex materials to black holes \cite{Polkovnikow2011Colloquium,Eisert2015Quantum,Hayden2007Black}. 
In isolated quantum systems, non-equilibrium dynamics typically lead to thermalization, where local observables evolve towards thermal equilibrium values, erasing the memory of their initial states. This paradigm is formalized by the eigenstate thermalization hypothesis (ETH)
~\cite{Deutsch1991Quantum,Srednicki1994Chaos,Rigol2008Thermalization}. However, mechanisms that defy thermalization and ETH do exist. Prominent examples include integrability \cite{Sutherland2004beautiful}, many-body localization \cite{Nandkishore2015many,Abanin2019Colloquium}, quantum many-body scars ~\cite{Bernien2017Probing,Bluvstein2021Controlling,Turner2018weak}, and Hilbert space fragmentation \cite{Sala2020Ergodicity,Adler2024Observation}. 

Thermalization-breaking phenomena also emerge in driven quantum many-body systems. Unlike their static counterparts, these systems lack energy conservation, typically leading to unbounded heating and the eventual decay of all non-trivial correlations~\cite{Lazarides2014Periodic,Lazarides2014Equilibrium,Alessio2014Long}. Strategies to circumvent heating include introducing strong spatial disorder, which can induce Floquet many-body localization~\cite{Ponte2015ManyBody,Lazarides2015Fate,Khemani2016Phase,Else2016Floquet,Yao2017Discrete,guo2023observation}, 
or operating in the high-frequency driving regime, where energy absorption is parametrically suppressed, giving rise to an exponentially long-lived prethermal state before the systems' eventual thermalization~\cite{Abanin2015Exponentially,Kuwahara2016Floquet,Abanin2017Rigorous,Else2017Prethermal,Else2020Long,Luitz2020Prethermalization}.
Recent theoretical works suggest that heating can also be suppressed in the intermediate-frequency regime through dynamical freezing~\cite{Haldar2021Dynamical,Haldar2024Dynamical}, where generic, interacting, and disorder-free systems can exhibit non-ergodic behaviors under strong periodic drives. In this scenario, breaking of ergodicity and thermalization stems from the emergence of new approximate (stroboscopic) conservation laws---absent in the undriven quantum chaotic system---rather than disorder-induced localization or high-frequency-induced prethermalization. Such emergent conservation laws fracture the Hilbert space into dynamically disconnected sectors, giving rise to almost perfect freezing of the emergent conserved quantities for arbitrary initial states. While many-body dynamical freezing  has been predicted in theory~\cite{Haldar2021Dynamical,Haldar2024Dynamical,Guo2025Dynamical,Mukherjee2024Floquet}, experimental observation of such a distinct and peculiar phenomenon remains elusive to date. 


Here, we present the first experimental observation of dynamical freezing in a strongly interacting ensemble of approximately $10^4$ spins associated with nitrogen-vacancy (NV) centers in diamond (Fig.~\ref{fig:fig1}).
Despite inherent on-site disorder and random spin-spin couplings, we observe robust preservation of total spin magnetization along the NV axis at specific ``freezing points", where the drive detuning matches even-integer multiples of the driving frequency (Figs.~\ref{fig:fig1}$\textbf{d}$ and \ref{fig:fig1}$\textbf{e}$). Remarkably, this emergent conserved quantity persists over up to $200$ driving cycles, exceeding the interaction-limited coherence time $T_2$ by more than an order of magnitude. We also observe coherent oscillatory micromotions, in excellent agreement with our theoretical predictions and numerical simulations. 
Perturbing the system away from freezing conditions restores Floquet thermalization (Fig.~\ref{fig:fig1}{\bf e}). Harnessing this effect, we develop a novel ac magnetometry protocol that combines dynamical freezing with periodic dynamical decoupling (PDD). 
Unlike conventional PDD schemes limited by $T_2$, our approach extends the optimal sensing window far beyond this coherence barrier while maintaining frequency selectivity. 
Under the same experimental conditions, we achieve a $4.3$ dB sensitivity improvement over conventional PDD methods. 
Our results establish dynamical freezing as both a fundamental thermalization-breaking mechanism in non-equilibrium quantum matter and a practical protocol for enhancing quantum sensing with broad applications across condensed matter~\cite{schirhagl2014nitrogen,casola2018probing,rovny2024nanoscale} and biology~\cite{aslam2023quantum,du2024single}.

\vspace{.5cm}
\noindent\textbf{\large{}Experiment setup and model Hamiltonian}{\large\par}
\noindent Our experiments are performed with a bulk diamond sample containing a dense ensemble of NV centers, represented by blue spins in Fig.\,\ref{fig:fig1}{\bf a}. Each NV center exhibits a triplet spin ground-state manifold, where two states  $\ket{m_s=0}\equiv\ket{\uparrow}$ and $\ket{m_s=-1}\equiv\ket{\downarrow}$ are used in this work. By using green lasers, these states are distinguished via spin-dependent florescence detection and each spin is initialized in $\ket{\uparrow}$ \cite{doherty2013nitrogen}. 
A static external magnetic field $B_0 = 432 ~\mathrm{G}$ aligns with one of the four diamond crystallographic axes (denoted as the $z$ axis), endowing this NV group with a distinct spin transition frequency from the other three orientations. These NVs are coherently manipulated by a near-resonant, global microwave field with programmable amplitude, frequency, and phase. 

In the rotating frame, the effective Hamiltonian governing the interacting NV spin ensemble can be described by ($\hbar=1$)
\begin{equation}
\begin{aligned}
H(t) = H_0 + \Omega \hat{S}_x +  h_z(t) \hat{S}_z.\end{aligned}
\label{Eq:System_Hamiltonian}
\end{equation}
Here, $H_0=\sum_{i<j} J_{ij}( \hat{s}_x^i \hat{s}_x^j + \hat{s}_y^i \hat{s}_y^j - \hat{s}_z^i \hat{s}_z^j ) + \sum_i h_i \hat{s}_z^i$
describes the static Hamiltonian.  $\hat{S}_{\mu}=\sum_{i} \hat{s}_{\mu}^i$ denotes the total spin operator, with $\hat{s}_{\mu}^{i}=\hat{\sigma}_\mu^i/2$ ($\mu \in \{x,y,z\}$) and $\hat{\sigma}_\mu^i$ being the Pauli matrices for the $i$-th spin.  $J_{ij}=J_0(3\cos^2\theta_{ij}-1)/r_{ij}^3$ characterizes spin-spin coupling strength, where $J_0 = 2 \pi \times 52 \text{ MHz} \cdot \text{nm}^3$, $r_{ij}$ and $\theta_{ij}$ are the distance and relative polar angle between the $i$-th and $j$-th spins. Disordered on-site fields $h_i$ are largely suppressed by dynamical decoupling \cite{de2010universal}. 
The Rabi frequency $\Omega$  and detuning $h_z(t)$ of the spin transition are controlled by global microwave fields.

The Floquet driving scheme for investigating dynamical freezing is illustrated by the pulse sequence in Fig.\,\ref{fig:fig1}\textbf{a}. 
After initializing each spin to the state $\ket{\psi}$, the system evolves under Floquet driving with period $T$ (Floquet driving frequency $\omega = 2\pi/T$). The native Hamiltonian alternates between $H_1= H_0+\Omega \hat{S}_x + h_z\hat{S}_z$ and $H_2= H_0+\Omega \hat{S}_x - h_z\hat{S}_z$ every half Floquet cycle. We consider a strong-amplitude, intermediate-frequency Floquet driving scenario with $h_z > \omega > \Omega, |J_{ij}|$. We analyze the system dynamics using a strong driving Magnus expansion in a moving frame~\cite{Haldar2021Dynamical} and derive the effective Floquet Hamiltonian $H_F=\sum_{n=0}^\infty H_F^{(n)}$, where $H_F^{(0)} =  H_0 + \frac{2\Omega}{h_z T} \left[(\sin \frac{h_z T}{2}) \hat{S}_x - (1-\cos\frac{h_z T}{2})\hat{S}_y \right]$ (Methods). The first-order term $H_F^{(1)}$ vanishes due to the reflection symmetry of the driving sequence. 
%
As a result, $H_F$ manifests an emergent conserved quantity $\langle\hat{S}_z\rangle$ that fractures the Hilbert space (Fig.\,\ref{fig:fig1}\textbf{b})  up to the second order, when the following freezing condition is satisfied:
\begin{equation}
h_z T = 2\pi \times 2k, \quad k= \pm1,\pm2,\cdots.
\label{Eq:freezingCondition}
\end{equation}  
At freezing points, the spin ensemble becomes ``frozen" under Floquet driving and its $\langle\hat{S}_z\rangle$ preserves. The leading symmetry-breaking term $\frac{\Omega^3}{4h_z^2} \hat{S}_x$ in $H_F^{(2)}$ governs the relaxation time at early evolution, while the late-time dynamics can be characterized by the diagonal ensemble average of Floquet eigenstates~\cite{Haldar2021Dynamical} (see Methods and Supplementary Information Sec.~I).
  

\vspace{.5cm}
\noindent\textbf{\large{}Observation of dynamical freezing}{\large\par}
\noindent We implement the Floquet driving Hamiltonian $H(t)$ by using the pulse sequence shown in Fig.\,\ref{fig:fig1}\textbf{c}. We start with preparing each spin in a target state $\ket{\psi(\theta,\varphi)}$ through optical pumping followed by two microwave rotations, where $\theta$ and $\varphi$ respectively denote the polar and azimuthal angles in the Bloch sphere. We then evolve the system under a Floquet driving with period $T=4\,\mathrm{\mu s}$ ($\omega= 2\pi \times 0.25$\,MHz). To suppress on-site disorders, a dynamical decoupling sequence is embedded in the Floquet cycle, where fast, equally spaced $\pi$-pulses 
are inserted in the Floquet driving pulses with alternating detunings $\pm h_z$ (Fig.~\ref{fig:fig1}\textbf{c}). 
After $N$ cycles, the total magnetization $\langle \hat{S}_{\mu}\rangle~(\mu\in\{x,y,z\})$  is measured by applying an optional $\pi/2$-pulse followed by fluorescence detection. 


Figure\,\ref{fig:fig1}\textbf{d} shows the time-dependent $z$-magnetization for varying $h_z$ with initial state being $\ket{\psi(0,0)}=\ket{\uparrow}$. 
When $h_z = 0$ the magnetization coherently oscillates at the Rabi frequency $\Omega = 2\pi\times 99.04(5) ~\mathrm{kHz}$ along with an amplitude decay primarily due to the spin-spin interactions. 
As $h_z$ increases, the $z$-magnetization oscillates more slowly and freezes at $h_z/(2\pi) = 0.5\,\mathrm{MHz}$, which is the first freezing point according to Eq.~\eqref{Eq:freezingCondition}. Further increasing $h_z$, the $z$-magnetization freezes only at discrete detunings, precisely matching the theoretical predictions.  
This behavior is better visualized by computing cumulative time averaged data, $ (\int_{0}^{t_f}\langle \hat{S}_z(t')\rangle_c dt')/t_f$. Data points for various $t_f$ are plotted against $h_zT$ in Fig.~\ref{fig:fig1}\textbf{e}. Three distinct peaks occur at the predicted freezing points $h_z T/(2\pi) = 2,4,6$ with heights close to unity for all $t_f$, providing unambiguous experimental evidence of dynamical freezing in our interacting spin ensemble.



 \begin{figure*}[tp!]
\includegraphics[width=0.75\linewidth]{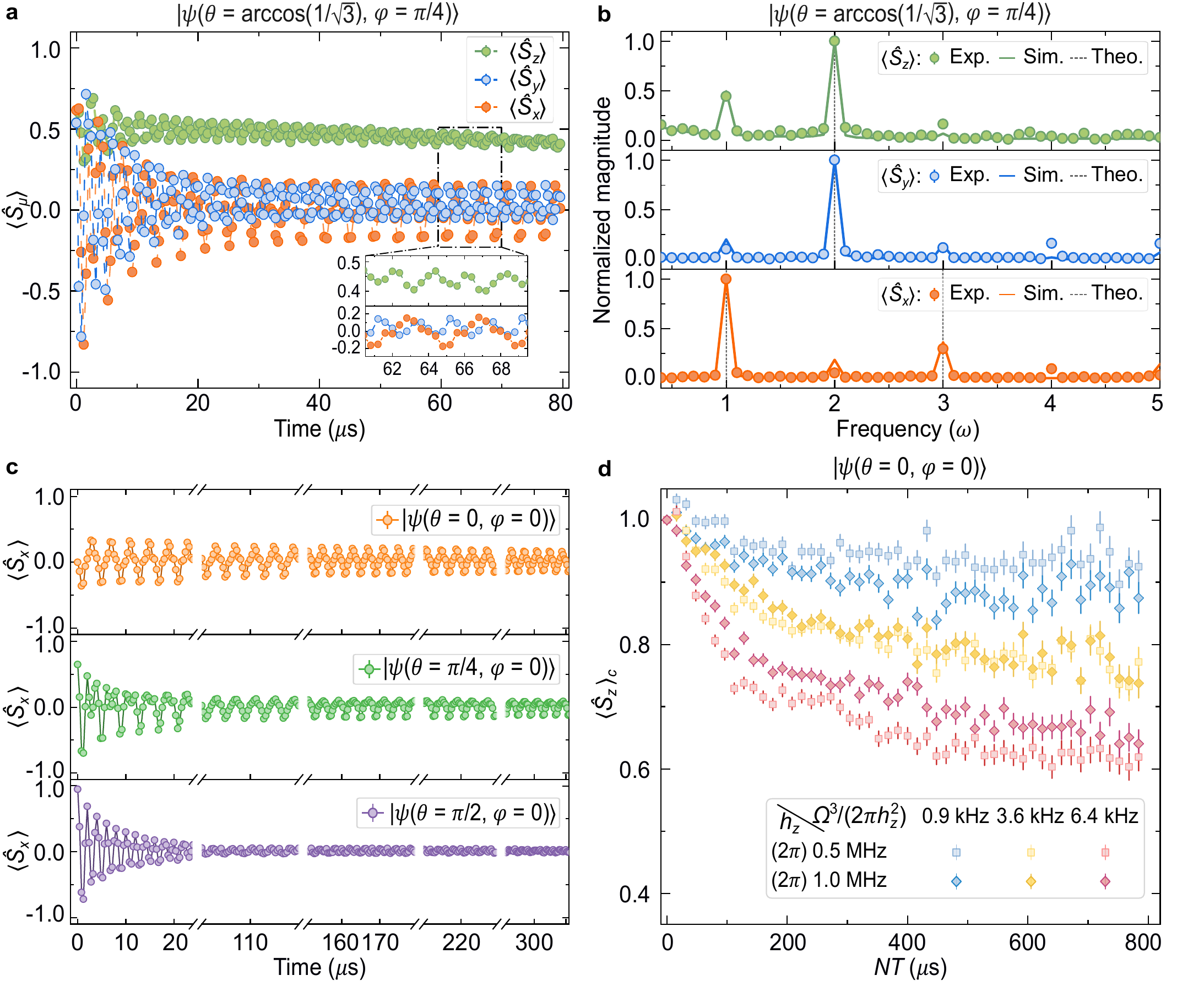}
\caption{\textbf{Micromotions and long-time behaviors at dynamical freezing points.}
\textbf{a}, Spin magnetizations $\langle \hat{S}_{\mu} \rangle \ (\mu\in\{x,y,z\})$ at the first freezing point $[h_z/(2\pi)=0.5\,\text{MHz}$, $\Omega/(2\pi)=0.1\,\mathrm{MHz}]$ are measured every $T/10=0.4\,\mathrm{\mu s}$, for an initial state with equal projection onto the $x,y,z$ axes. Inset: zoomed-in micromotion oscillations. 
\textbf{b}, Experimental (data points) and simulated (solid lines) micromotion spectra are obtained by Fourier transforming their temporal data between 40 and 80\,$\mathrm{\mu s}$. Their dominant frequency peaks precisely align with theoretical predictions of kick operators (vertical dashed lines). 
\textbf{c}, Time evolution of $\langle \hat{S}_x\rangle$ for three initial states at the first freezing point with $\Omega/(2\pi) = 0.1\ \text{MHz}$. The $\langle \hat{S}_x\rangle$ oscillation exhibits an initial decay and stabilizing to an amplitude which depends on the polar angle $\theta$ of the initial state. The stable section persists well beyond the coherence time $T_2=10.2(5)\,\mu$s.  
\textbf{d}, Plots of $\langle \hat{S}_z \rangle_c$ versus evolution time $NT$ at the first and second freezing points with the initial state $|\psi(0,0)\rangle$.  $\langle \hat{S}_z \rangle_c$ exhibits a relatively fast initial decrease and a subsequent slow decay. Data points with the same $\Omega^{3}/h_z^2$ are presented in similar colors and display similar line shapes.  
Data points are connected by lines in \textbf{a} and \textbf{c}, serving as a guide to the eye. 
Experiment is repeated over $2\times10^{4}$ trials for data points in \textbf{a-c} and $8\times10^{3}$ trials in \textbf{d}, with error bars representing one standard deviation.}
\label{fig:fig2fig3}
\end{figure*}

The intermediate driving frequency allows us to resolve the fine-grained intra-period micromotion dynamics, and gain deeper insight into the system's non-equilibrium evolution at freezing points.
Specifically, 
we measure the spin magnetization with a time step of $T/10$ for an initial state $\ket{\psi(\arccos(1/\sqrt{3}), \pi/4)}$, for which $\langle \hat{S}_x\rangle = \langle \hat{S}_y\rangle = \langle \hat{S}_z\rangle = 1/\sqrt{3}$. Fig.~\ref{fig:fig2fig3}\textbf{a} shows the temporal evolution of $\langle\hat{S}_{\mu}\rangle\  (\mu\in\{x,y,z\})$ for the first freezing point (see Extended Data Fig.~\ref{ExtendDataFigMicromotion} for results at the second freezing point). 
From this figure, the mean value of $\langle \hat{S}_z\rangle$ maintains finite for a long time due to the emergent $U(1)$ conservation law, whereas those for $\langle \hat{S}_x\rangle$ and $\langle \hat{S}_y\rangle$ decay to zero rapidly due to the lack of such conservation. Yet, all the three magnetization components exhibit coherent oscillations, with the oscillation amplitude for $\langle \hat{S}_x\rangle$ larger than that of $\langle \hat{S}_y\rangle$, which in turn is larger than that of $\langle \hat{S}_z\rangle$. 
%
We further investigate micromotion frequencies by performing Fourier transformation to the time-domain data in the $[40, 80]\,\mathrm{\mu s}$ window (excluding the initial decay). As shown in Fig.~\ref{fig:fig2fig3}\textbf{b}, the spectra of $\langle \hat{S}_z\rangle$ and $\langle \hat{S}_y\rangle$ exhibit a single dominant peak at $h_z=2\omega$. However, the spectrum of $\langle \hat{S}_x\rangle$ reveals two dominant peaks at $\omega$ and $3\omega$. 
To understand these dynamics, we analytically compute Floquet kick operators by the Floquet-Magnus expansion~\cite{Bukov2015Universal} in the moving frame (Methods). These observed peaks precisely match the theoretical predictions, as indicated by vertical dashed lines. We also carry out numerical simulations using the discrete truncated Wigner approximation method~\cite{Schachenmayer2015Many}, which agree well with the experiment observations.

Figure~\ref{fig:fig2fig3}\textbf{c} shows the micromotion dynamics of $\langle \hat{S}_x\rangle$ across extended time windows at the first freezing point for three initial states. 
For the initial state $\ket{\psi(0,0)} = \ket{\uparrow}$ (top panel), the $\langle \hat{S}_x\rangle$ oscillation persists throughout the entire scanning range with minimal amplitude decay, despite the latest time of 300\,$\mu$s exceeding the interaction-limited coherence time $T_2=10.2(5)\,\mu$s by over an order of magnitude. The coherence is extended due to the emergent conservation law: the system periodically returns to the initial state $\ket{\psi(0,0)}=\ket{\uparrow}$, which is the only state in the subspace with $\langle \hat{S}_z\rangle=1$, as illustrated in Fig.~\ref{fig:fig1}\textbf{b}.   
For $\ket{\psi(\pi/4,0)}$ in a larger subspace (middle panel, Fig.~\ref{fig:fig2fig3}\textbf{c}), transitions between states within this subspace are still allowed under Floquet driving even at freezing points. As a result, $\langle \hat{S}_x \rangle$ exhibits an initial decay within the timescale of $T_2$ and stabilizes to a reduced amplitude. Similar behaviors are seen with $\ket{\psi(\pi/2,0)}$ (bottom panel), whose equilibrium amplitude nearly vanishes as the subspace with $\langle \hat{S}_z\rangle=0$ contains the most states. 

We further measure $\langle \hat{S}_z\rangle$ at $NT$ with an initial state $\ket{\psi(0,0)}$ to study the dependence of freezing dynamics on driving parameters. As shown in Fig.~\ref{fig:fig2fig3}\textbf{d}, the resulting $\langle \hat{S}_z\rangle$ curves exhibit two-stage dynamics, consisting of an initial fast relaxation followed by a slow decay. Notably, data sharing the same value of $\Omega^{3}/h_z^2$---the coefficient of the leading symmetry-breaking term in $H_F^{(2)}$---display similar lineshapes despite substantially different driving parameters.  
These results confirm that the early-stage dynamics at freezing points are governed by this coefficient and imply its control over the late-time behaviors as well. We attribute the second-stage slow decay to the residual on-site disordered fields (Supplementary Information Fig.~S3).

\begin{figure*}[tp!]
\includegraphics[width=1.0\linewidth]{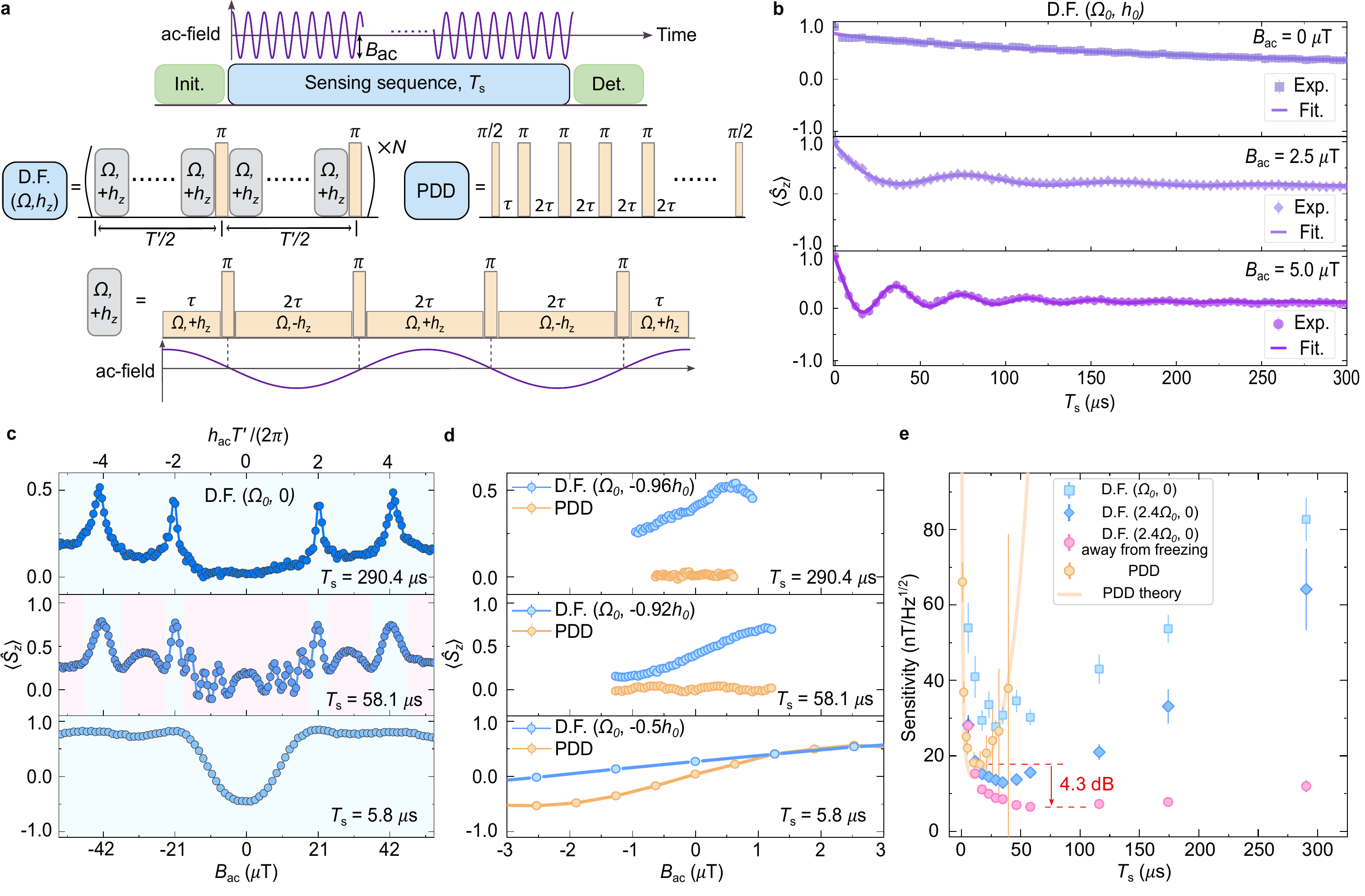}
\caption{\textbf{Dynamical-freezing enhanced magnetometry.} 
\textbf{a}, Pulse sequences for ac magnetic field sensing. NV electron spins serving as sensors are initialized with green lasers, manipulated by a microwave sequence (blue box) during which spins interact with an applied ac field of amplitude $B_{ac}$ (purple wavy line), and read out via fluorescence detection. Two sensing protocols are employed: dynamical freezing (D.F., middle-left sequence) and periodic dynamical decoupling (PDD, middle-right sequence). 
The D.F. sensing sequence is derived from the sequence in Fig.~\ref{fig:fig1}\textbf{c}. 
Both protocols use an equally spaced $\pi$-pulse train (inter-pulse interval  $2\tau=0.1\,\mu s$) synchronized to the field, acting as a frequency filter. 
\textbf{b}, Temporal response of $\langle\hat{S}_z\rangle$ for different $B_{ac}$ at the first freezing point. Data points are fit to an decaying sinusoidal function. 
\textbf{c}, $\langle\hat{S}_z\rangle$ versus $B_{\mathrm{ac}}$ at fixed $T_s$ using the D.F. sequence with $\Omega = \Omega_0$ and $h_z = 0$. 
At late time (top panel), the curve is analogous to the freezing spectrum in Fig.~\ref{fig:fig1}\textbf{e}, with distinct freezing points at certain $B_{ac}$, causing a field-induced detuning $h_{ac}$ satisfying $h_{ac}T^{\prime}=4\pi k$ ($k$ is a non-zero integer). Intermediate and early time behaviors (middle and bottom panels) exhibit coherent oscillations, and steep slopes are observed at places away from freezing points (pink regions in middle panel). 
\textbf{d}, Comparison between D.F. (blue) and PDD (orange) sensing protocols. PDD shows steeper slopes at $T_s = 5.8~\mathrm{\mu s}<T_2$ (lower panel), while for $T_s$ exceeding $ T_2$ (upper and middle panels), the D.F. protocol achieves significantly larger slopes than PDD.  
\textbf{e}, Sensitivity of the D.F. and PDD sensing protocols. D.F. optimal sensitivities near freezing points (e.g., blue regions in \textbf{c}) are presented as blue squares for $\Omega=\Omega_0$ and blue diamonds for $\Omega=2.4\Omega_0$. Data points for regions away from freezing points (e.g., pink regions in \textbf{c}) are displayed as pink circles. PDD experimental sensitivities (orange circles) are compared to theory (orange line). The lowest attainable sensitivity using D.F. is approximately 4.3\,dB lower than PDD.  
Experimental data points are averaged over $8\times10^3$ repetitions, with error bars representing one standard deviation. $\Omega_0/(2\pi)=0.1\,$MHz, $h_0/(2\pi)=0.5$\,MHz. 
}\label{fig:fig4}
\end{figure*}

\vspace{.5cm}
\noindent\textbf{\large{}Dynamical-freezing enhanced magnetometry}{\large\par}
\noindent The sharp spectral features and long lifetimes in dynamical freezing phenomena reveal pronounced potential for enhancing quantum metrology. 
When the system resides on a steep slope of the magnetization lineshape (e.g. near dashed lines in Fig.~\ref{fig:fig1}\textbf{e}), adding even a small magnetic field along the $z$-axis alters the effective detuning, resulting in drastic change in spin magnetization. 
%
Based on this principle, we devise a dynamical-freezing-enhanced sensing protocol using the pulse sequence shown in Fig.~\ref{fig:fig4}\textbf{a}. We initialize all spins in $\ket{\uparrow}$, implement a sensing sequence (blue box) while the spins interact with an applied ac field (purple curve), and then perform fluorescence detection to extract the information about the applied field. 
The devised protocol (middle-left panel) adapts the dynamical freezing sequence in Fig.~\ref{fig:fig1}\textbf{c} by inserting two $\pi$-pulses at $T/2$ and $T$ within each Floquet period. To preserve dynamical freezing, the detuning sign of the Floquet drives is inverted for the second half of each period.  
For suppressing local disordered fields, this sequence incorporates equally spaced $\pi$-pulses with an inter-pulse time of 2$\tau$, which acts as a frequency filter enabling narrow bandwidth ac magnetometry and noise spectroscopy~\cite{Degen2017RMP,pham2012enhanced,zhao2012sensing}. 
The $\pi$-pulse phases follow \{$x, x, -x, -x$\} for mitigating rotation errors. 
As a proof-of-principle demonstration, we apply an ac field with an amplitude of $B_{ac}$ oscillating at 3.78\,MHz, matching the filter center frequency. As shown in the lower panel of Fig.~\ref{fig:fig4}\textbf{a}, the applied field is synchronized with, and thus rectified by, the $\pi$-pulse train. As the driving detuning in the modified sequence is always positive, the rectified ac field introduces additional detuning to the Floquet driving system. 
We characterize this method and compare its sensing performance with conventional ac magnetometry using a PDD sequence (middle-right panel in Fig.~\ref{fig:fig4}\textbf{a}). Our experimental results are shown in Fig.~\ref{fig:fig4}\textbf{b-e}. 

Figure~\ref{fig:fig4}\textbf{b} displays the temporal response to an applied field when the Floquet system operates at the first freezing point, where $h_z=2\omega$, $\Omega/(2\pi)=\Omega_0/(2\pi)=0.1$\,MHz. When no field is applied ($B_{ac}=0$, upper panel), $\langle \hat{S}_z \rangle$ exhibits a slow decay primarily due to pulse errors, consistent with the expected dynamical freezing behavior. 
When a weak field is introduced ($B_{ac}=2.5\,\mu$T, middle panel), $\langle \hat{S}_z \rangle$ exhibits a decaying oscillation, substantially deviating from the initial value at early times. 
Raising amplitude to $B_{ac}=5\,\mu$T (lower panel) increases both amplitude and frequency of the $\langle \hat{S}_z \rangle$ oscillation, indicating a stronger response. 
Notably, these coherent oscillations caused by applied magnetic fields persist well beyond $T_2$ due to dynamical freezing. 

To separate the field-induced response from Floquet detuning, we set the Floquet drive on resonance ($h_z=0$) and measure $\langle \hat{S}_z \rangle$  for various $B_{ac}$ at sensing times $T_s$. 
In the late-time regime ($T_s=290.4\,\mu$s), where temporal oscillations of $\langle \hat{S}_z \rangle$ are nearly fully damped, the resulting $\langle \hat{S}_z \rangle$ versus $B_{ac}$ (top panel, Fig.~\ref{fig:fig4}\textbf{c}) exhibits reflection symmetry with respect to $B_{ac}=0$. The curve for $B_{ac}>0$ and the mirrored curve for $B_{ac}<0$ show line shapes similar to those in Fig.~\ref{fig:fig1}\textbf{e}, with distinct freezing points emerging at certain $B_{ac}$. These points match a modified freezing condition $h_{ac} T^{\prime}= 4\pi k$ (top horizontal axis, $k$ is a non-zero integer). $h_{ac} = \frac{2}{\pi}\gamma_{NV}B_{ac}$ quantifies the field-induced detuning, and  $T^{\prime}$ contains $\pi$-pulse durations during which the external field is on. 
At an intermediate time $T_s=58.1\,\mu$s (middle panel, Fig.~\ref{fig:fig4}\textbf{c}), $\langle \hat{S}_z \rangle$ exhibits coherent oscillations across the entire scan range despite the sensing time significantly exceeding $T_2$. In contrast to data near freezing points (blue regions), rapid oscillations also appear at regions away from freezing points (pink regions).  
In the early-time regime ($T_s=5.8\,\mu$s, lower panel in Fig.~\ref{fig:fig4}\textbf{c}), the oscillations are less pronounced due to insufficient time for the sensors' response to accumulate.

Next, we compare our protocol with the standard PDD method with results shown in Fig.~\ref{fig:fig4}\textbf{d}.  
We sweep the field amplitude $B_{ac}$ and examine the sensor response under both sensing sequences at different times. At a short time $T_s = 5.8\,\mu s$, the PDD method produces a sinusoidal $\langle \hat{S}_z \rangle$ with maximal slope at $B_{ac}=0$ (orange circles, lower panel). For comparison, we set $h_z=-0.5h_0$ for the dynamical freezing protocol so that the first freezing point is shifted and its slope is centered at $B_{ac}=0$ (blue circles). In this regime, the PDD method exhibits a steeper gradient. However, the advantage shifts dramatically at longer timescales. 
When $T_s=58.1\;\mathrm{\mu s}$, both methods develop faster oscillations due to extended phase accumulation (middle panel). The PDD method suffers from substantial reduction due to interaction-induced decoherence, while the dynamical freezing protocol retains a signal amplitude comparable to its short-time response. This difference is more evident at a longer time $T_s=290.4\,\mu$s (upper panel), where the PDD signal nearly vanishes while the dynamical freezing protocol maintains a robust response.

To further benchmark sensing performance, we compute the experimental sensitivities (Methods) for both protocols and present the data in Fig.~\ref{fig:fig4}\textbf{e}. 
The sensitivity of the dynamical freezing protocol is extracted from the experimental $\langle \hat{S}_z \rangle$ versus $B_{ac}$ curves, e.g., those in Fig.~\ref{fig:fig4}\textbf{c}. The optimal sensitivities near the freezing points (denoted by blue symbols in Fig. ~\ref{fig:fig4}\textbf{e}) are determined by the largest slope within the blue-shaded regions (see, for example, the middle panel of Fig.~\ref{fig:fig4}\textbf{c}).
In addition, we extract the optimal sensitivities away from the freezing points (pink regions) and plot these separately as pink circles. 
The experimental PDD sensitivities (orange circles) serve as a benchmark, plotted along with the theoretically predicted values (orange solid line, see Methods). 
%
As $T_s$ increases, the PDD sensitivity initially decreases, reaching a minimum of $18(3) ~\mathrm{nT/\sqrt{Hz}}$ near $T_2$. Beyond $T_2$, the sensitivity increases rapidly due to decoherence that degrades signal-to-noise ratio. 
In contrast, the dynamical freezing sensitivity with $\Omega=\Omega_0$ (blue squares) achieves its minimum of 28(2)$\,\mathrm{nT/\sqrt{Hz}}$, higher than the PDD's value, although this occurs at 29\,$\mu$s later than $T_2$. This larger sensitivity is due to small spectral slopes near the first freezing point at large $T_s$, as evidenced by the results in Extended Data Fig.~\ref{ExtendDataFigDFsensing}c. 
As predicted by Eq.~\eqref{Eq:sensing_response} in Methods, the slopes near freezing points increase with Rabi frequency $\Omega$. Raising Rabi frequency to $\Omega=2.4\Omega_0$ (blue diamonds) greatly improves the sensitivities, achieving a minimum value of $12.9(6) ~\mathrm{nT/\sqrt{Hz}}$ at $T_s=34.8\,\mu$s.     
When the Floquet driving system operates away from freezing points,  the sensitivities (pink circles) are even lower than those near the freezing points. Although the system is not frozen, experimental results indicate that the coherence times are substantially extended and sharper slopes are seen in the signal response, resulting in improved sensitivity. 
The optimal sensitivity achieved is $6.5(6) ~\mathrm{nT/\sqrt{Hz}}$, representing a 4.3\,dB improvement over the best PDD sensitivity. 
Notably, this optimum occurs at a sensing time later than the best time when operating near freezing points. Beyond this point, the sensitivity increases only marginally, possibly due to pulse errors. 
We also measure sensitivities for varying Floquet driving parameters and show the full datasets in Supplementary Information Figs.~S13-15.

\vspace{.5cm}
\noindent\textbf{\large{}Conclusion and outlook}{\large\par}

\noindent 
In summary, we have experimentally demonstrated dynamical freezing in a mesoscopic solid-state spin system, observing its characteristic signatures through persistent spin polarization and coherent micromotion. In contrast to previously reported breaking of thermalization and ergodicity, our observed dynamical freezing does not rely on disorder or high-frequency driving, but arises from emergent conservation laws and is independent of the initial state. In addition, by leveraging this unconventional dynamics, we have developed a dynamical-freezing-enhanced ac magnetometry protocol that shows substantial sensitivity enhancement over traditional techniques. 

Our results open up new avenues for future fundamental studies and practical applications. In particular, it would be interesting and important
to explore dynamical-freezing-based non-equilibrium phases, such as time crystals~\cite{Khemani2016Phase,Else2016Floquet,Yao2017Discrete} and Floquet symmetry-protected topological phases~\cite{Zhang2022Digital,Dumitrescu2022Dynamical}, with NV ensembles or other near-term
quantum platforms. For practical sensing, further sensitivity enhancements can be achieved through systematic analysis of Fisher information, suppressing pulse errors via advanced dynamical decoupling sequences, and reducing inhomogeneity of microwave and static magnetic fields. We also anticipate that the enhancement of this approach will be more pronounced in spin ensembles with higher density. 
Our protocol requires only global control, making it readily applicable to real-world sensing scenarios. This creates new opportunities to enhance magnetic field sensitivity without sacrificing spatial resolution for future studies in condensed matter physics, chemistry, and biology.

\bibliography{NVDF}

\newpage

\vspace{.5cm}
\noindent\textbf{\large{}Methods}{\large\par}

\setcounter{figure}{0} %
\renewcommand{\figurename}{Extended Data Fig.}

\vspace{.3cm}
\noindent\textbf{Micromotion dynamics}

\noindent By applying the Floquet-Magnus expansion~\cite{Bukov2015Universal} to the Hamiltonian in the moving frame $H_\text{mov} = H_0 + \Omega \left[\cos\theta(t) \hat{S}_x + \sin\theta(t) \hat{S}_y \right]$ with $\theta(t) = - h_z \int_0^t dt' \mathrm{sgn}(\sin\omega t') $, we obtain the leading-order Floquet kick operator governing the micromotion oscillations within each period, $K_F^{(1)}(t)=$
\begin{equation}
\frac{\Omega}{h_z} \times
\begin{cases} 
\sin( h_z t ) \hat{S}_x - (1-\cos h_z t) \hat{S}_y ,&t\in [0, T/2]\\ 
\sin( h_z t ) \hat{S}_x + (1-\cos h_z t) \hat{S}_y ,&t\in [T/2, T].
\end{cases}
\end{equation}
Here we have imposed the dynamical freezing condition Eq.~\eqref{Eq:freezingCondition}, which results in $K_F^{(1)}(0)=K_F^{(1)}(T)=K_F^{(1)}(T/2)=0$. 

After a sufficiently long evolution time ($NT \gg T_2$), $\langle \hat{S}_x(NT) \rangle $ and $\langle \hat{S}_y(NT)\rangle $ at discrete times $NT$ approach their equilibrium values of zero, due to thermalization dynamics determined by the stroboscopic effective Hamiltonian $H_F = H_0$. $\langle \hat{S}_z(NT) \rangle $ approximately preserves its initial value due to the emergent conservation law at freezing points. For inter-stroboscopic times $t\in[NT, (N+1)T]$ (with $t^* = t- NT$), the micromotion oscillations controlled by the leading-order kick operator $K^{(1)}_{F}$ 
are characterized by
\begin{align} \label{Eq:micromotion}
\langle \hat{S}_x(t)\rangle &\approx \frac{\Omega}{h_z}\langle \hat{S}_z(NT)\rangle \times
\begin{cases} 
1 - \cos h_z t^*& t^* \in [0, T/2] \\
- \left(1 - \cos h_z t^*\right) & t^* \in [T/2, T],
\end{cases} \nonumber \\
\langle \hat{S_y}(t)\rangle &\approx -\frac{\Omega}{h_z} \langle \hat{S_z}(NT)\rangle \times \sin h_z t^*, \\
\langle \hat{S_z}(t)\rangle &\approx \langle \hat{S_z}(NT)\rangle \times \left[1 - \frac{\Omega^2}{h_z^2} \left(1 - \cos h_z t^*\right) \right]. \nonumber 
\end{align}
Eq.~\eqref{Eq:micromotion} predicts the oscillation amplitudes and dominant frequencies of the spin magnetization in different axes, which are compared to experimental and numerical data in Fig.~\ref{fig:fig2fig3}\textbf{b} and Extended Data Fig.~\ref{ExtendDataFigMicromotion}\textbf{b}.  We also observe certain minor frequency peaks in both experimental results and numerical simulations, which are likely attributed to high-order kick operators. 

\vspace{.3cm}
\noindent\textbf{Numerical methods}

\noindent We employ the exact diagonalization method to compute the eigenstates of the single-period Floquet evolution unitary $U(T,0)$, satisfying $U(T,0)\ket{\mu_n} = e^{-i\mu_n }\ket{\mu_n}$ where $\mu_n$ and $\ket{\mu_n}$ denote the $n$th eigenvalue and eigenstate. In the infinite-time limit, the expectation value for an observable $\hat{O}$ is characterized by the diagonal ensemble average over these Floquet eigenstates~\cite{Haldar2021Dynamical}: $\lim_{t\to \infty}\langle \psi(t)| \hat{O} | \psi(t)\rangle = \sum_n |c_n|^2  \langle \mu_n | \hat{O} | \mu_n \rangle = \langle O\rangle_{DE}$, where $\ket{\psi(0)} = \sum_n c_n \ket{\mu_n}$ and the subscript ''DE" represents diagonal ensemble. As shown in Supplementary Information Sec.~I.~B, we utilize this diagonal ensemble average to characterize the long-time behaviors of dynamical freezing in small-size systems. This analysis also corresponds to the ultimate fate of the second-stage slow decay in Fig.~\ref{fig:fig2fig3}\textbf{d}.

For our mesoscopic NV ensemble (approximately $10^4$ spins with long-range interactions), we employ the discrete truncated Wigner approximation (DTWA)~\cite{Schachenmayer2015Many} to simulate the Floquet driving dynamics at intermediate evolution times. 
DTWA is a semiclassical  Monte Carlo method that samples the Wigner function in discrete phase space, capturing the dynamics of one- and two-point correlations at the mean-field level~\cite{Schachenmayer2015Many}. 
The discrete phase space of a spin-$1/2$ system consists of four phase points $\alpha \in \{(0,0), (0,1), (1,0), (1,1)\}$. For each phase point, we define the phase point operator 
$A_\alpha = (1 + \vec{r}_\alpha \cdot \vec{\sigma}) / 2$, where $\vec{r}_{(0,0)} = (1,1,1),\vec{r}_{(0,1)} = (-1,-1,1) ,\vec{r}_{(1,0)} = (1,-1,-1),\vec{r}_{(1,1)} = (-1,1,-1)$, and $\vec{\sigma} = (\sigma_x, \sigma_y, \sigma_z)$ consists of spin-$1/2$ Pauli matrices. 
The Weyl symbol $O^W_\alpha = \mathrm{Tr}(\hat{O} A_\alpha) / 2$ maps a quantum operator $\hat{O}$ to the four-point phase space. For a given density matrix $\rho$, the discrete Wigner function is the corresponding Weyl symbol of $\rho$, $w_\alpha = \mathrm{Tr}(\rho A_\alpha) / 2$. 
In the Heisenberg picture, the Wigner function is always fixed at its initial value. The evolution of an operator $\hat{O}$ can be semi-classically computed by 
$\langle \hat{O}(t) \rangle = \sum_\alpha w_\alpha(0) O^W_\alpha(t) \approx \sum_\alpha w_\alpha(0) O^{W,cl}_\alpha(t)$, where
$O^{W,cl}_\alpha(t)$ is the classically evolved Weyl symbol. Therefore, we conduct Monte Carlo sampling over the initial discrete Wigner function, then keep track of the observable evolution of each trajectory, and finally average over the trajectory ensemble to obtain $\langle \hat{O}(t) \rangle$.
The classical equations of motion for spin observables are given by the Poisson brackets, and we numerically calculate their evolution through solving differential equations by the Runge-Kutta method.

\vspace{.3cm}

\vspace{.3cm}
\noindent \textbf{Experimental data analysis} 

\begin{figure*}[tp!]
\includegraphics[width=0.8\linewidth]{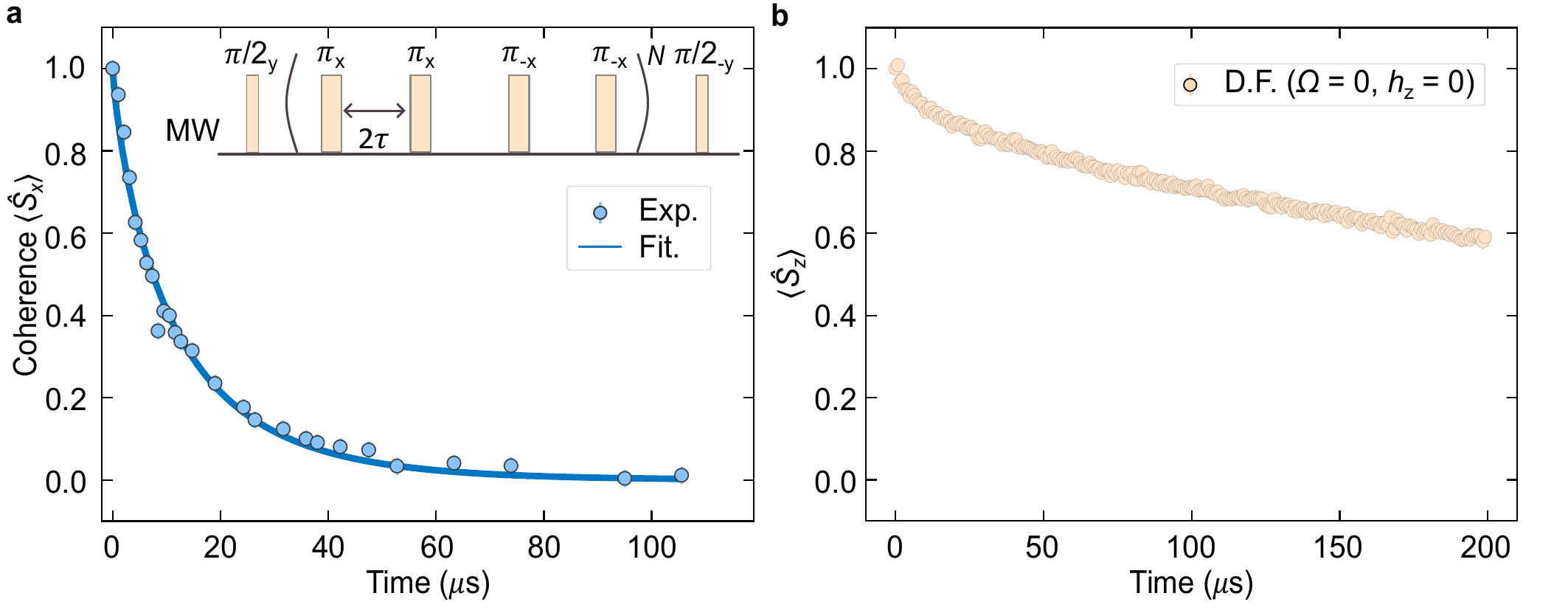}
\caption{\textbf{Characterization of coherence time and pulse errors.}
\textbf{a,} NV electron spin coherence under dynamical decoupling with a $\pi$-pulse duration of $32\,\mathrm{ns}$ and inter-pulse spacing $2\tau = 0.1\,\mathrm{\mu s}$. Experimental data points of $\langle\hat{S}_x \rangle$ (blue circles with error bars) are fit to a stretched exponential decay function $e^{-(t/T_2)^\alpha}$ (blue line), yielding $T_2 = 10.2(5)\,\mathrm{\mu s}$ and $\alpha= 0.81(5)$. Inset: dynamical decoupling pulse sequence using resonant microwave fields. 
\textbf{b,} Measured spin magnetization $\langle\hat{S}_z\rangle$ under the dynamical freezing sequence with $\Omega=0$. The signal exhibits a slow decay mainly due to $\pi$-pulse errors. These data points serves as references and are used to correct pulse errors for data in Fig. \ref{fig:fig1}\textbf{d} and Fig.~\ref{fig:fig2fig3}\textbf{d}. 
}\label{fig:ExtendDataFigCoherence}
\end{figure*}

\begin{figure*}[tp!]
\includegraphics[width=0.9\linewidth]{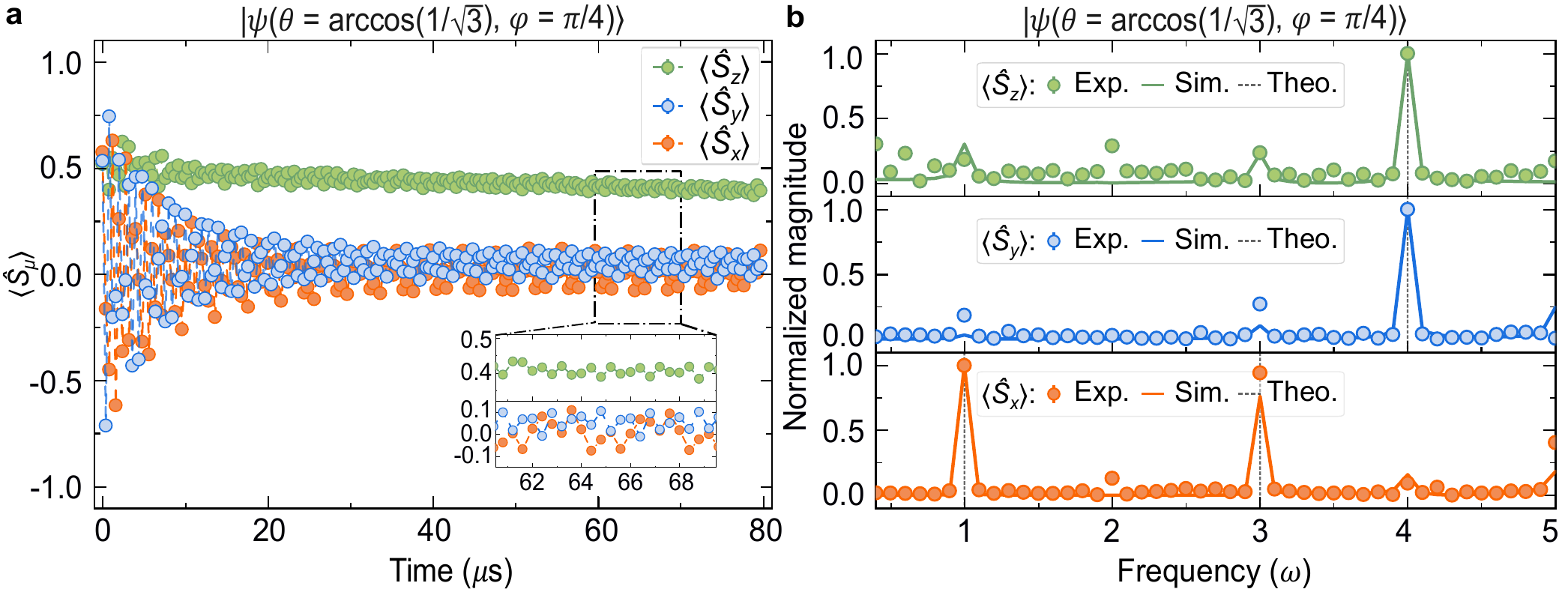}
\caption{\textbf{Micromotions at the second freezing point.}
\textbf{a,} The temporal evolution of spin magnetization along the $x,y,z$ axes at the second dynamical freezing point with the detuning $h_z/(2\pi)=1.0\;\mathrm{MHz}$ and Rabi frequency $\Omega/(2\pi)=0.1\;\mathrm{MHz}$. 
\textbf{b,} Experimental (dots) and numerically simulated (solid lines) spectra extracted by Fourier transforming the temporal data of spin magnetization in the time window of $[40, 80]\,\mathrm{\mu s}$. Their dominant frequency peaks are consistent with theoretical predictions of kick operators (vertical dashed lines). Each data point with one standard deviation is obtained through $2\times10^4$ trials. 
}
\label{ExtendDataFigMicromotion}
\end{figure*}

\begin{figure*}[tp!]
\includegraphics[width=1.0\linewidth]{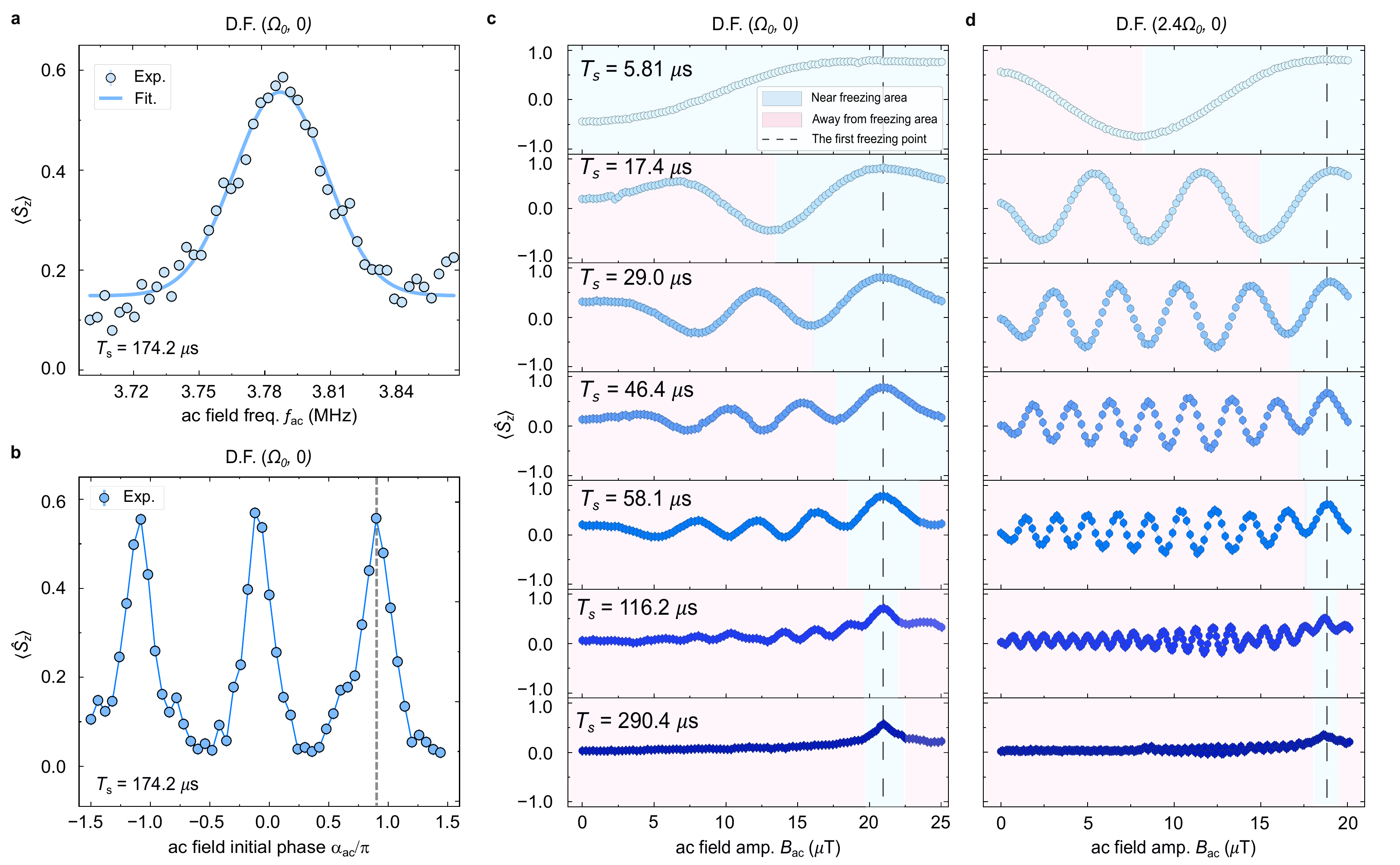}
\caption{\textbf{Dynamical-freezing sensing of external ac fields.} \textbf{a,} Plots of $\langle \hat{S}_{z}\rangle$ as a function of the ac field frequency $f_{ac}$ at a fixed sensing time $T_{s}=174.24~\mathrm{\mu s}$. These data points are fit to a Gaussian function (solid blue line), yielding the fitted center frequency $f_{ac,\mathrm{exp}}=3.7870(7)~\mathrm{MHz}$. \textbf{b,} Plots of $\langle \hat{S}_{z}\rangle$ versus the initial phase $\alpha_{ac}$ at a fixed sensing time $T_s = 174.24~\mathrm{\mu s}$.  A Rabi frequency of $\Omega_0/(2\pi)=0.1\,\mathrm{MHz}$ is used in \textbf{a} and \textbf{b}. \textbf{c, d,}  $\langle \hat{S}_{z}\rangle$ responses to the ac field amplitude $B_{ac}$ under various $T_{s}$ with $h_z=0$ and different Rabi frequencies, $\Omega=\Omega_0$ in \textbf{c} and $\Omega=2.4\Omega_0$ in \textbf{d}. Black dashed line indicates the first freezing point. Regions near the freezing point (blue shaded areas) and away from the freezing point (pink shaded areas) are highlighted and their sensitivities are analyzed separately. Each data point and error bar represents the average value and one standard deviation over $8\times10^3$ repetitions.
}\label{ExtendDataFigDFsensing}
\end{figure*}

\noindent The total spin magnetization, $\langle \hat{S}_{\mu}\rangle\  (\mu\in \{x,y,z\})$, is obtained by differential measurements using two sequential and nearly identical experiments. In one experiment, a $\pi$-pulse is applied before fluorescence detection, flipping spin states along the $\mu$-axis. Fluorescence count difference between the two experiments is normalized to yield $\langle \hat{S}_{\mu}\rangle$.  This treatment mitigates experimental imperfections such as laser amplitude fluctuations.  
To isolate the intrinsic dynamical freezing phenomenon, we exclude pulse errors in dynamical decoupling sequence.  
We first measure the spin magnetization dynamics using $\Omega=0$ as references. The results are shown in Extended Data Fig. \ref{fig:ExtendDataFigCoherence}\textbf{b}. Data points in Fig. \ref{fig:fig1}\textbf{d} and Fig.~\ref{fig:fig2fig3}\textbf{d} with the initial state of $\ket{\psi(0,0)}$ are normalized relative to this reference, following the relation: $\langle \hat{S}_{\mu} (t)\rangle_c = \langle \hat{S}_{\mu} (t)\rangle/\langle \hat{S}_{z} (t,\Omega=0)\rangle$.


\vspace{.3cm}
\noindent\textbf{Dynamical-freezing sensing of ac magnetic fields}

\noindent 
The Hamiltonian of NV electron spins under the external ac field can be written as 
\begin{equation}
     H_{ac}(t) = H  + \gamma_{\mathrm{NV}}B(t)\hat{S}_{z},
 \label{H:Sensing}
\end{equation}
where $H$ is the system Hamiltonian in the absence of ac fields [Eq.~\eqref{Eq:System_Hamiltonian}], $\gamma_{\mathrm{NV}}=2\pi\times 28.03\;\mathrm{GHz/T}$ denotes the gyromagnetic ratio of the NV electron spins, and $B(t)=B_{ac}\cos(2\pi f_{ac}t+\alpha_{ac})$. 

The embedded dynamical decoupling sequence acts as a frequency filter with the center frequency $f_{ac,\mathrm{theo}}=1/(4\tau+2t_{\pi})=3.7878\;\mathrm{MHz}$, where $2\tau$ and $t_{\pi}$ denote the pulse spacing and $\pi$ pulse duration, respectively. To validate this, we experimentally sweep the ac field frequency at a fixed amplitude $B_{ac}$ and a sensing time ($T_s=174.24\;\mathrm{\mu s}$). Gaussian fit to the experimental spectrum (Extended Data Fig. \ref{ExtendDataFigDFsensing}\textbf{a}) yields a central frequency $f_{ac,\mathrm{exp}}=3.7870(7)\;\mathrm{MHz}$, consistent with the theoretical prediction. 
Then, we set the applied field at $f_{ac,\mathrm{exp}}$ and scan $\alpha_{ac}$ from  $-1.5\pi$ to $1.5\pi$ under the dynamical freezing sequence with zero detuning ($h_z=0$) to characterize the phase dependence. Extended Data Fig. \ref{ExtendDataFigDFsensing}\textbf{b} shows the measured $\langle S_{z}\rangle$ as a function of $\alpha_{ac}$, from which the phase $\alpha_{ac}=0.9\pi$ for maximal response is identified (indicated by dashed line). 

Using the resonant frequency and phase for maximum response, we derive the zeroth-order effective Hamiltonian under small deviations from the freezing condition $h_z=2\gamma_\text{NV}B_{ac}/\pi+2k\omega$:
\begin{equation}
H_F^{(0)} \approx H_0 + \frac{\Omega}{h_z} \frac{2\gamma_\text{NV}B_{ac}}{\pi} \hat{S}_x.
\label{Eq:sensing_response}
\end{equation}
According to Eq.~\eqref{Eq:sensing_response}, the response to the fields is stronger near the first freezing peak ($h_z =\pm2\omega$), compared to higher-order peaks, which can be evidenced from Fig. \ref{fig:fig4}\textbf{c}. 

We experimentally vary the Rabi frequency $\Omega /( 2\pi)$ from 0.1\,$\mathrm{MHz}$ to 0.28\,$\mathrm{MHz}$ while fixing the detuning at $h_z=0$. Representative experimental results for $\Omega / (2\pi) = 0.1\;\mathrm{MHz}$ and $\Omega/ (2\pi) =0.24\;\mathrm{MHz}$ are presented in Extended Data Fig.~\ref{ExtendDataFigDFsensing}\textbf{c} and \textbf{d}. Measurements span from $B_{ac} = 0$ to an amplitude across the first freezing point. All experimental datasets show progressive sharpening for freezing peaks with extended sensing time, from top to bottom in Extended Data Fig.~\ref{ExtendDataFigDFsensing}\textbf{c} and \textbf{d}. However, the peak amplitude diminishes due to accumulated pulse errors. Oscillatory features also occur at regions away from the freezing point where slopes steeper than those near the freezing peaks are seen. These oscillations are still significant even at $T_s \gg T_2$ and become faster at higher Rabi frequencies. We extract the experimental sensitivities and separate the data into two parts: near the freezing points (blue shaded areas in Extended Data Fig.~\ref{ExtendDataFigDFsensing}\textbf{c} and \textbf{d}), and away from freezing points (pink shaded areas in Extended Data Fig.~\ref{ExtendDataFigDFsensing}\textbf{c} and \textbf{d}).
Future theoretical works will investigate the detailed relationship between Floquet driving parameters, coherence times, and slopes away from freezing points.

\vspace{.3cm}
\noindent \textbf{Experimental sensitivity} 

\noindent Experimental sensitivity for measuring ac magnetic field amplitudes is derived using~\cite{Degen2017RMP,Barry2020RMP}  
\begin{equation}\label{Eq:sensitivity}
    \eta = \frac{\sigma_s(N)}{\left|\frac{d\langle\hat{S}_z\rangle}{dB_{ac}}\right|}\sqrt{T_{t}},
\end{equation}
where $T_t$  and $\sigma_s(N)$ respectively denote the total experimental integration time and the standard deviation of the measured  $\langle\hat{S}_z\rangle$ of $N$ experimental trials.  $\frac{d\langle\hat{S}_z\rangle}{dB_{ac}}$ is the slope of the response $\langle \hat{S}_z \rangle$ as a function of $B_{ac}$. 
Experimental $\frac{d\langle\hat{S}_z\rangle}{dB_{ac}}$ is obtained from the sinusoidal fits to experimental data for the PDD method. For the dynamical freezing method, smoothing over three consecutive data points (moving-average) is applied for mitigating statistical fluctuations prior to differentiation.

\vspace{.3cm}
\noindent \textbf{Theoretical sensitivity of the PDD method}

\noindent The sensitivity of the PDD-based ac magnetometry can be formulated as~\cite{Barry2020RMP}
\begin{equation}
    \eta_{\mathrm{theo,PDD}} = \frac{\pi}{2\gamma_{\mathrm{NV}}}\frac{e^{(T_s/T_2)^\alpha}}{C}\frac{\sqrt{T_{s}+T_{\mathrm{o}}}}{T_s},
\end{equation}
where $T_s$ is the sensing time, $T_2$ represents the coherence time under the PDD sequence, $\alpha$ denotes the stretching exponent of the decoherence profile, $C$ is the overall readout efficiency, and $T_{o}=T_{I}+T_{R}+T_{d}$ defines the total overhead time, comprising spin initialization ($T_{I}$), optical readout ($T_{R}$) and necessary delay time ($T_{d}$). 

In this experiment, the parameters $T_2 = 10.2(5)\;\mathrm{\mu s}$ and $\alpha = 0.81(5)$ are determined from the NV coherence dynamics (see Extended Data Fig. \ref{fig:ExtendDataFigCoherence}\textbf{a}). The experimental times include $T_{I} = 40\;\mathrm{\mu s}$, $T_{R}= 2\;\mathrm{\mu s}$, and $T_{d} = 5.133\;\mathrm{\mu s}$. The readout efficiency $C=0.0139$ of the NV ensemble (approximately 10$^{4}$ NVs) is derived via $C=1/(\sigma_s(N)\sqrt{N})$ with standard deviation $\sigma_s(N)$ of $N$ experimental trials.

\vspace{.6cm}
\noindent \textbf{\large{}Data availability}
The data presented in the figures and that support  other findings of this study will be publicly available at Zenodo.org. 

\ 

\vspace{.6cm}
\noindent\textbf{\large{}Code availability}
The data analysis and numerical simulation codes will be publicly available at Zenodo.org.

\vspace{.5cm}
\noindent\textbf{\large{}Acknowledgement} We thank Xun Gao, Zhe-Xuan Gong, Alexey V. Gorshkov, Guanghui He, Hsin-Yuan Huang, Wen Wei Ho, Dayou Yang, Bingtian Ye, and  Shunyao Zhang for helpful discussions.  
This work was supported by the Innovation Program for Quantum Science and Technology (grant nos. 2021ZD0302203, 2021ZD0301601 and 2021ZD0301605), the National Natural Science Foundation of China (grant nos. T2225008, 12075128, and 123B2072), the Tsinghua University Dushi Program, the Shanghai Qi Zhi Institute (grant nos. SQZ202317, and SQZ202318), the Tsinghua University Initiative Scientific Research Program, and the Ministry of Education of China. L.-M. D. acknowledges in addition support from the New Cornerstone Science Foundation through the New Cornerstone Investigator Program. P.-Y. H. acknowledges the start-up fund from Tsinghua University. 

\vspace{.5cm}
\noindent\textbf{\large{}Author contributions} Y.-N. L., D. Y., and P.-Y. H. designed the experiments. P.-Y. H. conceived the sensing idea. Y.-N. L. carried out the experiments under the supervision of P.-Y. H. and L.-M. D.. D. Y. and Y. M. performed the numerical simulations under the supervision of H.-Z. Z. and D.-L. D.. P.-Y. H., Y.-N. L., D. Y., Y. M., S. J., X.-Q. M., Y.-J. X., X.-Y. C., Y.-Q. L., H. Z., D.-L. D., and L.-M.D. analyzed the data and conducted the theoretical analysis. All authors contributed to the experimental set-up, the discussions of the results, and the writing of the manuscript.

\end{document}